# Algorithms for the Iterative Estimation of Discrete-Valued Sparse Vectors

Susanne Sparrer, Robert F.H. Fischer
Institute of Communications Engineering, Ulm University, 89081 Ulm, Germany
Email: susanne.sparrer@uni-ulm.de, robert.fischer@uni-ulm.de

*Abstract*—In Compressed Sensing, a real-valued sparse vector has to be estimated from an underdetermined system of linear equations. In many applications, however, the elements of the sparse vector are drawn from a finite set. For the estimation of these discrete-valued vectors, matched algorithms are required which take the additional knowledge of the discrete nature into account. In this paper, the estimation problem is treated from a communications engineering point of view. A powerful new algorithm incorporating techniques known from digital communications and information theory is derived. For comparison, Turbo Compressed Sensing is adapted to the discrete setup and a simplified and generalized notation is presented. The performance of the algorithms is covered by numerical simulations.

## I. INTRODUCTION

In a number of communication scenarios, the noisy receive vector $y \in \mathbb{R}^{K \times 1}$ at one time instance is given by[1]

$$y = Ax + n ,  \quad (1)$$

where the transmitted vector $x \in \mathbb{R}^{L \times 1}$ is sparse, i.e., only $s \ll K < L$ elements are non-zero. Throughout this paper, the non-zero elements are assumed to be drawn from the finite set $\mathcal{C} = \{\pm 1\}$, which corresponds to the transmission of 2ASK modulated signals. The measurement matrix $A \in \mathbb{R}^{K \times L}$ corresponds to the channel matrix in a communication scenario. The measurements are corrupted by noise $n$ which is assumed to be i.i.d. zero-mean Gaussian distributed with variance $\sigma_n^2$ per component.

Since $K \ll L$, a *discrete-valued sparse* vector has to be estimated from an *underdetermined* set of linear equations at the receiver. If the exact sparsity is known, which is assumed throughout this paper, the problem which has to be solved is given by ($\mathcal{C}_0 \stackrel{\text{def}}{=} \mathcal{C} \cup \{0\}$)

$$\hat{x} = \underset{\tilde{x} \in \mathcal{C}_0^L}{\operatorname{argmin}} \|y - A\tilde{x}\|_2^2 \quad \text{s.t.} \quad \|\tilde{x}\|_0 = s . \quad (2)$$

The problem of the estimation of a *discrete-valued sparse* vector from an *underdetermined* system of linear equations appears in various fields of digital communications, e.g., in sensor networks, where $L$ sensors with very low activity independently transmit binary data and a fusion center with $K$ antennas has to reconstruct which centers were active and which data has been transmitted by them [1]. There are also many other applications like peak-to-average power reduction in orthogonal frequency-division multiplexing [2], the detection of pulse-width-modulated signals in radar applications [3], <u>co</u>de-book <u>e</u>xcited <u>l</u>inear <u>p</u>rediction (CELP) source coding [4], and Compressed Sensing-based cryptography [5].

In the literature, there is a vast amount of algorithms solving the standard <u>C</u>ompressed <u>S</u>ensing (CS) problem [6], where the sparse vector is assumed to be real-valued, in particular <u>O</u>rthogonal <u>M</u>atching <u>P</u>ursuit (OMP) [7], <u>I</u>terative <u>H</u>ard <u>T</u>hresholding (IHT) [8], and <u>I</u>terative <u>S</u>oft <u>T</u>hresholding (IST) [9], to mention the most prominent ones.

In *discrete compressed sensing*, however, the additional information that the elements of the estimated vector are from a finite set has to be taken into account adequately. Note that the estimation of a discrete-valued vector has combinatorial complexity in general. Furthermore, even if $x$ was real-valued, problem (2) would be nonconvex due to the sparsity constraint. In the literature, some algorithms have been proposed which solve a relaxed $\ell_1$-based, but still nonconvex, problem. These algorithms are extensions of the simplex algorithm and have a prohibitively high computational complexity [10].

During the last few years, some proposals on how to solve the original problem (2) have been made. The most obvious solution is the concatenation of a standard CS algorithm with successive symbolwise quantization [11]. While this approach has the advantage that it does not put any restrictions on the CS algorithm to be used, it suffers from poor performance because the knowledge of $x$ is disregarded *inside* the reconstruction algorithm. The performance can be improved if, instead of symbolwise quantization after the CS algorithm, vector quantization is employed [11]. Therefore, lattice decoders like, e.g., the <u>S</u>phere <u>D</u>ecoder (SD) [12] can be used. Although this approach performs much better than the simple symbolwise quantization, the knowledge of the discreteness of $x$ is still not used within the CS algorithm and the SD has a high computational complexity which grows rapidly with the sparsity and the noise power.

Another approach discussed in the literature is model-based compressed sensing [13]. In the case of discrete CS, this corresponds to quantization *inside* the reconstruction algorithm. In [14], however, it has been shown that the reconstruction does not benefit from the incorporation of the quantization in the OMP, since the reliability information contained in the analog estimates is eliminated if hard decisions are taken. It was shown that a gain can be achieved if the knowledge of the discrete nature of the signal is included while still using reliability information [14]. Unfortunately, the estimation of the required parameters is quite troublesome. Another variant of OMP using a minimum mean-squared error estimator has

This work was supported by Deutsche Forschungsgemeinschaft (DFG) under grant FI 982/8-1.

[1]Notation: $\|\cdot\|_p$ denotes the $\ell_p$ norm. $A_{(l,m)}$ is the element in the $l^{\text{th}}$ row and $m^{\text{th}}$ column of $A$, and $K_{i,i}$ denotes the $i^{\text{th}}$ diagonal element of $K$. $A^{\mathsf{T}}$ and $A^{-1}$ denote the transpose and the inverse of $A$, respectively. $\operatorname{diag}(a)$ denotes a diagonal matrix of appropriate size with entries of the vector $a$ as diagonal elements. $\operatorname{diag}(A)$ denotes a diagonal matrix with the same diagonal elements as $A$. $I$ is the identity matrix. $\mathcal{Q}_\mathcal{C}(\cdot)$: element-wise quantization to a given alphabet $\mathcal{C}$. $\Pr\{\cdot\}$: probability; $\mathrm{E}\{\cdot\}$: element-wise expectation. $\operatorname{Var}\{\cdot\}$: Variance. $\delta(\cdot)$: Dirac delta distribution.

been introduced in [15].

The Compressed Sensing problem has also been tackled from a channel coding perspective, leading to, e.g., the approximate message-passing (AMP) algorithm [16], [17] which is a modification of the message-passing algorithm which is used for the decoding of low-density parity-check codes [18]. In [16], a generalization of AMP has been proposed which handles cases with known a-priori distribution of the sparse vector. This algorithm, which is often denoted as Bayesian AMP (BAMP) or generalized AMP (GAMP), can be easily adapted to the discrete scenario discussed in this paper. There are also other approaches for (discrete) Compressed Sensing which transfer knowledge from channel coding to CS in terms of optimized measurement matrices, and adapted recovery algorithms for these special cases, cf. e.g., [19], [20]. Note that, however, these approaches are limited to special applications in which the measurement matrix can be chosen.

In this paper, the decoding problem is investigated from a digital communications and information theory perspective. Using the knowledge known from these fields, a powerful new algorithm is derived. Furthermore, the principle of Turbo Signal Recovery (TSR) [21] is revisited, adapted to the setting at hand, and presented in a simplified way.

The paper is organized as follows. In Sec. II, knowledge from communications engineering is utilized to build a new algorithm, which combines two basic principles of signal estimation. In Sec. III, Turbo Signal Recovery is adapted to discrete compressed sensing and a generalized version is presented. The performance of the algorithms is evaluated in Sec. IV, followed by brief conclusions in Sec. V.

## II. THE ITERATIVE MMSE-SF-ALGORITHM (IMS/Q)

In this section, a new algorithm is introduced. It is based on two main concepts which are both very well known in communications engineering. Remember that problem (2), which has to be solved, has two constraints: First, the estimation error in terms of squared Euclidean distance has to be minimized. Second, the sparsity condition has to be fulfilled. Furthermore, the elements of the estimated vector have to be from the finite set $\mathcal{C}_0$.

The new algorithm finds solutions for one of the constraints in an alternating fashion. In the first part of the algorithm, an estimate with respect to the _minimum mean-squared error (MMSE) criterion_ is calculated, in particular the first criterion is fulfilled via linear MMSE estimation. This approach has already been introduced to Compressed Sensing in the combination with the OMP in [15]. The restriction of the sparsity and of the finite alphabet is taken into account in the second step, where the so-called _soft values_ are computed. This approach is denominated as _soft feedback_ (SF). In communications engineering, it is used in many applications, e.g., for successive interference cancellation (SIC), a.k.a., decision-feedback equalization (DFE), in multiuser detection [22], [23], [24]. It has been applied to Compressed Sensing in the OMP in [14]. In the following, the two parts of the algorithm are derived.

### A. MMSE Estimation

For the derivation of the MMSE-based estimation, we assume the channel model (1), i.e., $y = Ax + n$. Furthermore, a previous estimate $\hat{x}$ of $x$ (e.g., the result from the previous iteration) is assumed to be given. Taking this knowledge into account, we calculate a new estimate $\tilde{x}$ which minimizes the squared Euclidean distance to $x$. The knowledge on the sparsity and on the discreteness of $x$ is ignored in this step. Following the standard approach to linear estimation, e.g., [25], the estimate is given by

$$\tilde{x} = \hat{x} + \Phi_{xx} A^\mathsf{T} \left( A \Phi_{xx} A^\mathsf{T} + \sigma_n^2 I \right)^{-1} (y - A\hat{x}) \ , \quad (3)$$

thereby $\Phi_{xx}$ is the covariance matrix of $x$. Note that $\hat{x}$ can be written as noisy variant of $x$, with

$$\hat{x} = x + d \ , \quad (4)$$

where $d$ is the error vector. This leads to the covariance matrix

$$\begin{aligned}\Phi_{xx} &= \mathrm{E}\{(x - \hat{x})(x - \hat{x})^\mathsf{T}\} \\ &= \mathrm{E}\{dd^\mathsf{T}\} = \Phi_{dd} \ ,\end{aligned} \quad (5)$$

where we assume that $\hat{x} = \mathrm{E}\{x\}$. Since the elements of $\hat{x}$ are (assumed to be) uncorrelated, $\Phi_{dd}$ is a diagonal matrix. Plugging in this into (3), it can be rewritten as

$$\tilde{x} = \hat{x} + \Phi_{dd} A^\mathsf{T} \left( A \Phi_{dd} A^\mathsf{T} + \sigma_n^2 I \right)^{-1} (y - A\hat{x}) \ , \quad (6)$$

where the part

$$B \stackrel{\mathrm{def}}{=} \Phi_{dd} A^\mathsf{T} \left( A \Phi_{dd} A^\mathsf{T} + \sigma_n^2 I \right)^{-1} \quad (7)$$

is denoted as equalization matrix.

When dealing with _discrete_ symbols, it is important that the estimate is not biased, i.e., that the diagonal elements of the end-to-end cascade $K = BA$ for the estimation of $x$ are equal to 1. Using $B$, given in (7), for equalization, $K$ is given by

$$K = \Phi_{dd} A^\mathsf{T} \left( A \Phi_{dd} A^\mathsf{T} + \sigma_n^2 I \right)^{-1} A \ . \quad (8)$$

Note that, due to the term $\sigma_n^2 I$, the diagonal elements of $K$ are always smaller than 1 and hence a bias is present which has to be removed before the decision of the symbols. To this end, a scaling matrix $W$ which compensates for the bias is introduced. It is given by

$$W = \mathbf{diag}(1/K_{1,1}, \ldots, 1/K_{L,L}) \ . \quad (9)$$

The _unbiased_ estimate calculates to

$$\tilde{x} = \hat{x} + W \Phi_{dd} A^\mathsf{T} \left( A \Phi_{dd} A^\mathsf{T} + \sigma_n^2 I \right)^{-1} \cdot (y - A\hat{x}) \ . \quad (10)$$

This estimate can be described as noisy variant of $x$, i.e.,

$$\tilde{x} = x + e \ , \quad (11)$$

with the additive error vector $e$ with variances $\sigma_{e,i}^2$ per component. The correlation matrix of the error $e = \tilde{x} - x$ calculates to

$$\begin{aligned}\Phi_{ee} &= \Phi_{dd} + W \Phi_{dd} A^\mathsf{T} \left( A \Phi_{dd} A^\mathsf{T} + \sigma_n^2 I \right)^{-1} A \Phi_{dd} W^\mathsf{T} \\ &\quad - W \Phi_{dd} A^\mathsf{T} \left( A \Phi_{dd} A^\mathsf{T} + \sigma_n^2 I \right)^{-1} A \Phi_{dd} \\ &\quad - \Phi_{dd}^\mathsf{T} A^\mathsf{T} \left( A \Phi_{dd} A^\mathsf{T} + \sigma_n^2 I \right)^{-1} A \Phi_{dd}^\mathsf{T} W^\mathsf{T} \\ &= \Phi_{dd} + W K \Phi_{dd} W^\mathsf{T} - W K \Phi_{dd} - \Phi_{dd}^\mathsf{T} K^\mathsf{T} W^\mathsf{T} \ .\end{aligned} \quad (12)$$

The calculation of the main diagonal elements $\sigma_{e,i}^2$, which

correspond to the error variance of the respective elements $\tilde{x}_i$, can be simplified to

$$\sigma_{e,i}^2 = [\boldsymbol{\Phi}_{ee}]_{(i,i)} = [\boldsymbol{\Phi}_{dd}]_{(i,i)} \cdot \frac{1 - K_{i,i}}{K_{i,i}} \ . \tag{13}$$

In the upper part of Fig. 1, the block diagram of IMS is shown, interpreting the algorithm as communication system. The middle part shows the end-to-end model, including the measurement matrix $\boldsymbol{A}$, the noise $\boldsymbol{n}$, and the linear MMSE estimation step. The second step of the algorithm uses the results ($\tilde{\boldsymbol{x}}$ and $\boldsymbol{\Phi}_{ee}$) of the first step. In the bottom part, the end-to-end model also including the second step is shown. The results of the second step ($\hat{\boldsymbol{x}}$ and $\boldsymbol{\Phi}_{dd}$) and $\boldsymbol{y}$ are used as input for the first step.

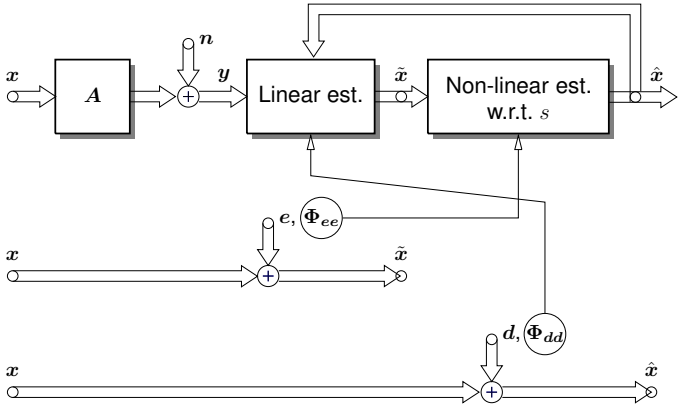

Fig. 1: Block diagram as communications system.

### B. Soft Feedback

In the first step of the new algorithm, the energy of the estimation error was minimized, thereby ignoring the sparsity constraint and the discreteness of $\boldsymbol{x}$. In the second step, based on the result from the first step, $\boldsymbol{x}$ is estimated with respect to the desired sparsity and the alphabet, now ignoring the constraint on the squared error. A similar step is also applied, e.g., in IHT and IST, where the non-sparse estimate from a first step is sparsified by the so-called thresholding step [8], [9]. Please note that, however, the thresholding in IHT and IST does not include any information on the probability density of the elements of $\boldsymbol{x}$.

The most obvious solution to include the sparsity and the restricted alphabet in the estimation of $\boldsymbol{x}$ would be hard quantization with respect to the alphabet, where the quantization threshold would be adapted such that the estimate matches the desired sparsity. The drawback of this approach is that there is no reliability information on the symbols after the quantization.

In the new algorithm proposed in this paper, another approach, the so-called *soft feedback* [26], is used. To this end, the expected value $\hat{x}_i$ of $x_i$ given the observation $\tilde{x}_i$ from the first step and given the channel model (11) and the probability density function $f_x(x)$ of $x$ is calculated. For the signal model at hand,

$$f_x(x) = \frac{s/2}{L}\delta(x+1) + \frac{L-s}{L}\delta(x) + \frac{s/2}{L}\delta(x-1) \ . \tag{14}$$

The soft values can be calculated by $\hat{x}_i = \mathcal{W}(\tilde{x}_i, \sigma_{e,i}^2, s)$ with

$$\mathcal{W}(\tilde{x}_i, \sigma_{e,i}^2, s) \stackrel{\text{def}}{=} \mathrm{E}\{x_i|\tilde{x}_i\}$$

$$= \frac{\frac{s}{2}\left(\mathrm{e}^{-\frac{(\tilde{x}_i-1)^2}{2\sigma_{e,i}^2}} - \mathrm{e}^{-\frac{(\tilde{x}_i+1)^2}{2\sigma_{e,i}^2}}\right)}{\frac{s}{2}\left(\mathrm{e}^{-\frac{(\tilde{x}_i-1)^2}{2\sigma_{e,i}^2}} + \mathrm{e}^{-\frac{(\tilde{x}_i+1)^2}{2\sigma_{e,i}^2}}\right) + (L-s)\cdot\mathrm{e}^{-\frac{\tilde{x}_i^2}{2\sigma_{e,i}^2}}}$$

$$= \frac{\sinh\left(\frac{\tilde{x}_i}{\sigma_{e,i}^2}\right)}{\cosh\left(\frac{\tilde{x}_i}{\sigma_{e,i}^2}\right) + \frac{L-s}{s}\cdot\mathrm{e}^{+\frac{1}{2\sigma_{e,i}^2}}} \ . \tag{15}$$

The variance of these symbols calculates as

$$\sigma_{d,i}^2 = \mathrm{Var}\{x_i|\tilde{x}_i\} = \mathrm{E}\{x_i^2|\tilde{x}_i\} - (\mathrm{E}\{x_i|\tilde{x}_i\})^2 \ , \tag{16}$$

where the expected value of $x_i^2$ given $\tilde{x}_i$ calculates to

$$\mathrm{E}\{x_i^2|\tilde{x}_i\} = \int_{-\infty}^{\infty} x^2 f_x(x_i|\tilde{x}_i)\,\mathrm{d}x$$

$$= \frac{\frac{s}{2}\left(\mathrm{e}^{-\frac{(\tilde{x}_i-1)^2}{2\sigma_{e,i}^2}} + \mathrm{e}^{-\frac{(\tilde{x}_i+1)^2}{2\sigma_{e,i}^2}}\right)}{\frac{s}{2}\left(\mathrm{e}^{-\frac{(\tilde{x}_i-1)^2}{2\sigma_{e,i}^2}} + \mathrm{e}^{-\frac{(\tilde{x}_i+1)^2}{2\sigma_{e,i}^2}}\right) + (L-s)\cdot\mathrm{e}^{-\frac{\tilde{x}_i^2}{2\sigma_{e,i}^2}}}$$

$$= \frac{\cosh\left(\frac{\tilde{x}_i}{\sigma_{e,i}^2}\right)}{\cosh\left(\frac{\tilde{x}_i}{\sigma_{e,i}^2}\right) + \frac{(L-s)}{s}\cdot\mathrm{e}^{+\frac{1}{2\sigma_{e,i}^2}}} \ . \tag{17}$$

Plugging in this to (16), yields the error variance as

$$\sigma_{d,i}^2 = \frac{\frac{L-s}{s}\cdot\mathrm{e}^{+\frac{1}{2\sigma_{e,i}^2}}\cdot\cosh\left(\frac{\tilde{x}_i}{\sigma_{e,i}^2}\right) + 1}{\left(\cosh\left(\frac{\tilde{x}_i}{\sigma_{e,i}^2}\right) + \frac{L-s}{s}\cdot\mathrm{e}^{+\frac{1}{2\sigma_{e,i}^2}}\right)^2} \ . \tag{18}$$

The end-to-end model, including the measurement matrix, the noise, and the estimations steps, again interpreted as communications system, is shown in Fig. 1, lower part (cf. (4)). In the next iteration of the algorithm, the MMSE estimation takes $\hat{\boldsymbol{x}}$ and $\boldsymbol{\Phi}_{dd} = \mathrm{diag}([\sigma_{d,1}^2, \ldots, \sigma_{d,L}^2])$ as input.

An example of the corresponding characteristic curve is given in Fig. 2 for $s/L = 0.1$ and $\sigma_{e,i}^2 \in \{0.01, 0.05, 0.5\}$. For comparison, the characteristic curve of hard quantization (black), as well as the ones applied in IHT ($\mathrm{T_H}(\tilde{x}_i, s)$, purple) and in IST ($\mathrm{T_S}(\tilde{x}_i, \tau)$, yellow), are also shown, where $\tau$ denotes a threshold to be optimized.

In the case of small error variance, i.e., very reliable prior estimates $\tilde{x}_i$, the curve of soft feedback (blue) tends to the one of hard quantization (black), i.e., hard decisions are made. If the error variance increases and thus the prior estimate $\tilde{x}_i$ becomes less reliable, however, the slope of the characteristic curve of soft thresholding decreases (green and red).

This algorithm represents an easily comprehensible solution to the problem of estimating a discrete-valued sparse vector from an underdetermined system of linear equations.

Since it iteratively performs, in an alternating fashion, MMSE estimation and soft feedback calculation, it is denoted by IMS/Q algorithm, where the trailing "Q" emphasizes the terminating quantization step which is required for the final result to be restricted to the alphabet. The pseudocode of the algorithm is given in Alg. 1. Note that, by simply adjusting the soft-feedback calculation and the quantization, this algorithm

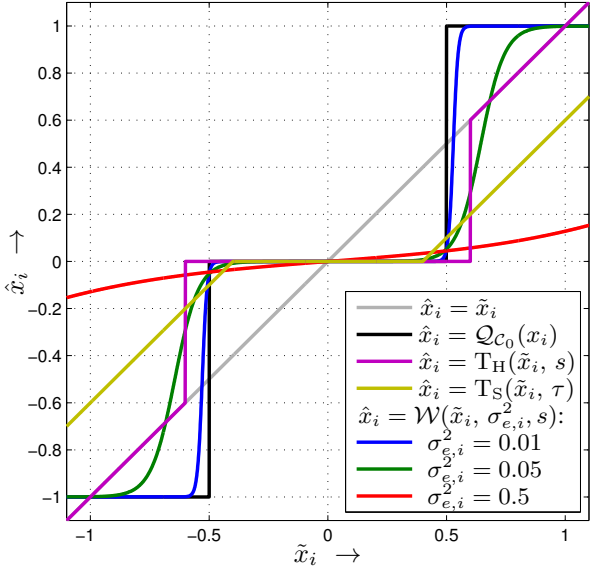

Fig. 2: Example of the characteristic curves for soft feedback ($s/L = 0.1$).

can be adapted to every desired set $\mathcal{C}$.

---

**Alg. 1** $\hat{\boldsymbol{x}} = \text{IMS/Q}\left(\boldsymbol{y}, \boldsymbol{A}, \sigma_n^2, s, \mathcal{C}_0\right)$

1: $\tilde{\boldsymbol{x}} = \boldsymbol{0}$, $\sigma_{d,i}^2 = s/L \ \forall \ i$
2: **while** stopping criterion not met {
   // Unbiased $\ell_2$-minimization
3: $\quad \tilde{\boldsymbol{x}} = \hat{\boldsymbol{x}} + \boldsymbol{W}\boldsymbol{\Phi}_{dd}\boldsymbol{A}^\mathsf{T}\left(\boldsymbol{A}\boldsymbol{\Phi}_{dd}\boldsymbol{A}^\mathsf{T} + \sigma_n^2 \boldsymbol{I}\right)^{-1}(\boldsymbol{y} - \boldsymbol{A}\hat{\boldsymbol{x}})$
4: $\quad \sigma_{e,i}^2 = [\boldsymbol{\Phi}_{dd}]_{(i,i)} \cdot (1 - K_{(i,i)})/K_{(i,i)}$
   // Minimization w.r.t. $s$ and $\mathcal{C}_0$ (for $\mathcal{C}_0 = \{-1, 0, +1\}$)
5: $\quad \hat{x}_i = \dfrac{\sinh\left(\frac{\tilde{x}_i}{\sigma_{e,i}^2}\right)}{\cosh\left(\frac{\tilde{x}_i}{\sigma_{e,i}^2}\right) + \frac{L-s}{s} \cdot e^{\frac{1}{2\sigma_{e,i}^2}}}$
6: $\quad \sigma_{d,i}^2 = \dfrac{\frac{L-s}{s} \cdot e^{\frac{1}{2\sigma_{e,i}^2}} \cdot \cosh\left(\frac{x_i}{\sigma_{e,i}^2}\right) + 1}{\left(\cosh\left(\frac{\tilde{x}_i}{\sigma_{e,i}^2}\right) + \frac{L-s}{s} \cdot e^{\frac{1}{2\sigma_{e,i}^2}}\right)^2}$
7: }
8: $\hat{\boldsymbol{x}} = \mathcal{Q}_{\mathcal{C}_0}(\hat{\boldsymbol{x}})$

---

This new algorithm can also be interpreted as advanced and adapted (to the discrete setup) variant of IHT/Q. In IHT/Q, the linear signal estimation in the first step is based on the correlation between the measurement matrix $\boldsymbol{A}$ and the residual $\boldsymbol{y} - \boldsymbol{A}\hat{\boldsymbol{x}}$, which can also be interpreted as the application of a matched filter in terms of communications engineering. In IMS/Q, this simple approach is replaced by the superior linear MMSE estimation. In the second step, the hard thresholding (or soft tresholding in the case of IST) is replaced by the calculation of the soft feedback, thereby including reliability information in terms of error variances as well as alphabet constraints.

## III. TURBO SIGNAL RECOVERY

In this section, the <u>T</u>urbo <u>S</u>ignal <u>R</u>ecovery (TSR) algorithm, proposed in [21], [27], is recapitulated and adapted to the discrete setup. It is shown that the algorithm can be significantly simplified without any change in performance. A comparison to IMS/Q is given.

Note that an adaptation of TSR to the dual problem, i.e., CS with *quantized measurements* $\boldsymbol{y}$ (instead of a discrete-valued sparse signal vector $\boldsymbol{x}$ which we assume) is introduced in [28].

While AMP [16], [17] is derived from message passing [18] which is often used for the decoding of LDPC codes, the main idea behind the TSR algorithm is to solve the standard compressed sensing problem with continuous-valued sparse vectors using another well-known approach from channel decoding, namely the turbo principle [29]. In contrast to the standard assumption common in the CS literature that the *column* vectors of $\boldsymbol{A}$ are normalized to unit length, the authors of the TSR algorithm require the *row* vectors of $\boldsymbol{A}$ to be normalized. Furthermore, the measurement matrix is expected to be constructed as random part of a unitary matrix $\boldsymbol{M}$ by $\boldsymbol{A} = \boldsymbol{S}\boldsymbol{M}$, where the selection matrix $\boldsymbol{S}$ is a random choice of the rows of an identity matrix of appropriate size. Furthermore, an auxiliary variable $\boldsymbol{z} = \boldsymbol{M}\boldsymbol{x}$ is introduced, which leads to $\boldsymbol{y} = \boldsymbol{S}\boldsymbol{z} + \boldsymbol{n}$. If $\boldsymbol{M}$ is a DCT (or DFT) matrix, $\boldsymbol{z}$ can be interpreted as frequency-domain representation of $\boldsymbol{x}$.

In order to generalize the TSR to a wider range of measurement matrices, we introduce a scaling matrix $\boldsymbol{C} = \text{diag}([c_1, \ldots, c_L])$. The measurement matrix is then given by $\boldsymbol{A} = \boldsymbol{S}\boldsymbol{M}\boldsymbol{C}$, and $\boldsymbol{z} = \boldsymbol{M}\boldsymbol{C}\boldsymbol{x}$. Note that the original construction is still included by $\boldsymbol{C} = \boldsymbol{I}$, and in the standard CS setup with normalized column vectors the scaling elements $c_i$ are given by $c_i^2 = 1/\sum_{j=1}^{K}((\boldsymbol{S}\boldsymbol{M})_{(j,i)})^2$.

The pseudocode of the TSR algorithm is shown in Alg. 2. As IMS/Q, the algorithm consists of two parts. First, an estimate on $\boldsymbol{z}$ is calculated, neglecting to sparsity constraint. In the second part, an estimate of $\boldsymbol{x}$ is calculated taking the sparsity constraint and the prior distribution of $x$ into account.

The first step is given by [21]
$$\boldsymbol{z}_\text{A}^\text{post} = \boldsymbol{z}_\text{A}^\text{pri} + \frac{\sigma_{\text{A,pri},z}^2}{\sigma_{\text{A,pri},z}^2 + \sigma_n^2} \boldsymbol{S}^\mathsf{T}(\boldsymbol{y} - \boldsymbol{S}\boldsymbol{z}_\text{A}^\text{pri}) \,, \qquad (19)$$

where $\boldsymbol{z}_\text{A}^\text{pri}$ is a prior knowledge on $\boldsymbol{z}$ (corresponding to the prior estimate of $\boldsymbol{x}$ used in IMS/Q) with *average* error variance $\sigma_{\text{A,pri},z}^2$. In standard TSR, the estimate $\boldsymbol{z}_\text{A}^\text{post}$ is later on transformed to an estimate of $\boldsymbol{x}$ which is required for the second step of the algorithm. In the following we show that the estimation of the artificial variable $\boldsymbol{z}$ is unnecessary and the estimation can be rewritten to work purely on $\boldsymbol{x}$. Plugging in $\boldsymbol{z} = \boldsymbol{M}\boldsymbol{C}\boldsymbol{x}$ into (19) leads to

$$\boldsymbol{M}\boldsymbol{C}\boldsymbol{x}_\text{A}^\text{post} = \boldsymbol{M}\boldsymbol{C}\boldsymbol{x}_\text{A}^\text{pri} + \frac{\sigma_{\text{A,pri},z}^2}{\sigma_{\text{A,pri},z}^2 + \sigma_n^2} \boldsymbol{S}^\mathsf{T}(\boldsymbol{y} - \boldsymbol{S}\boldsymbol{z}_\text{A}^\text{pri}) \,. \qquad (20)$$

Note that the average variance $\sigma_{\text{A,pri}}^2$ of $\boldsymbol{x}_\text{A}^\text{pri}$ differs from the average variance of $\boldsymbol{z}_\text{A}^\text{pri}$ and can be calculated by $\sigma_{\text{A,pri}}^2 = \sigma_{\text{A,pri},z}^2 / \bar{c}^2$, where $\bar{c}^2 = \frac{1}{L}\sum_{i=1}^{L} c_i^2$ is the average scaling factor from the rescaling of the measurement matrix. By left multiplying (20) with $(\boldsymbol{M}\boldsymbol{C})^{-1} = \boldsymbol{C}^{-1}\boldsymbol{M}^\mathsf{T}$ and with $\boldsymbol{S}\boldsymbol{M}\boldsymbol{C} = \boldsymbol{A}$, $\boldsymbol{x}_\text{A}^\text{post}$ can be directly calculated by

$$\boldsymbol{x}_\text{A}^\text{post} = \boldsymbol{x}_\text{A}^\text{pri} + \frac{\bar{c}^2 \sigma_{\text{A,pri}}^2}{\bar{c}^2 \sigma_{\text{A,pri}}^2 + \sigma_n^2} \cdot \boldsymbol{C}^{-1}\boldsymbol{M}^{-1}\boldsymbol{S}^\mathsf{T}(\boldsymbol{y} - \boldsymbol{A}\boldsymbol{x}_\text{A}^\text{pri}) \,.$$

Since $\boldsymbol{M}$ is unitary and with the non-normalized measurement matrix $U \stackrel{\text{def}}{=} SM$, which leads to $A = UC$, the calculation can be further simplified to ($\bar{c} I \approx C$)

$$\begin{aligned} x_A^{\text{post}} &= x_A^{\text{pri}} + \frac{\bar{c}^2 \sigma_{A,\text{pri}}^2}{\bar{c}^2 \sigma_{A,\text{pri}}^2 + \sigma_n^2} \cdot C^{-1} U^\top (y - A x_A^{\text{pri}}) \\ &= x_A^{\text{pri}} + \frac{\bar{c}^2 \sigma_{A,\text{pri}}^2}{\bar{c}^2 \sigma_{A,\text{pri}}^2 + \sigma_n^2} \cdot C^{-1} C^{-1} A^\top (y - A x_A^{\text{pri}}) \\ &\approx x_A^{\text{pri}} + \frac{\sigma_{A,\text{pri}}^2}{\bar{c}^2 \sigma_{A,\text{pri}}^2 + \sigma_n^2} \cdot A^\top (y - A x_A^{\text{pri}}) . \end{aligned} \quad (21)$$

Thus, $x_A^{\text{post}}$ can be directly estimated without the intermediate calculation of $z_A^{\text{post}}$. Note that the conversion was purely based on linear rearrangements of the equations, such that the final result is the same as the one of the original TSR algorithm in [21], with the additional generalization to scaled measurement matrices.

The same considerations apply to the calculation of the variance of the estimation error, cf. Line 4, Alg. 2. Note that, in contrast to IMS/Q, the TSR algorithm calculates *average* variances, the individual variances of the distinct elements are not taken into account.

As it is common for turbo decoders, the so-called extrinsic information (i.e., the information gained by this first step) is calculated and forwarded to the second part of the algorithm (cf. Lines 5 and 6, Alg. 2). In the second step of TSR, the soft values, which have also been discussed for IMS/Q, are calculated in a symbolwise fashion. As in the first step, the extrinsics of this step have to be calculated. They serve as input for the first step in the next iteration. It can be shown that for Gaussian signals unbiasing operation and extrinsic calculation coincide (cf. the extrinsic calculation [30]).

Since the final estimate has to be constrained to the alphabet, a final quantization step with respect to $\mathcal{C}_0$ is required. In case of known sparsity, which is assumed in this paper, the quantization threshold is adapted such that the estimate matches the desired sparsity. This TSR algorithm with final quantization is denoted by TSR/Q in the following.

---

**Alg. 2** $\hat{x} = \text{TSR/Q}(y, A, \sigma_n^2, s, \mathcal{C}_0)$

1: $x_A^{\text{pri}} = 0$, $\sigma_{A,\text{pri}}^2 = s/L$
2: **while** stopping criterion not met {
   // Estimation
3: $x_A^{\text{post}} = x_A^{\text{pri}} + \frac{\bar{c}^2 \sigma_{A,\text{pri}}^2}{\bar{c}^2 \sigma_{A,\text{pri}}^2 + \sigma_n^2} \cdot C^{-1} U^\top (y - A x_A^{\text{pri}})$
4: $\sigma_{A,\text{post}}^2 = \sigma_{A,\text{pri}}^2 - \frac{K}{L} \frac{(\bar{c}^2 \sigma_{A,\text{pri}}^2)^2}{\bar{c}^2 \sigma_{A,\text{pri}}^2 + \sigma_n^2}$
5: $\sigma_{B,\text{pri}}^2 = \sigma_{A,\text{ext}}^2 = \left( \frac{1}{\sigma_{A,\text{post}}^2} - \frac{1}{\sigma_{A,\text{pri}}^2} \right)^{-1}$
6: $x_B^{\text{pri}} = x_A^{\text{ext}} = \sigma_{A,\text{ext}}^2 \left( \frac{x_A^{\text{post}}}{\sigma_{A,\text{post}}^2} - \frac{x_A^{\text{pri}}}{\sigma_{A,\text{pri}}^2} \right)$
   // Soft feedback
7: $x_{B,i}^{\text{post}} = \mathrm{E}\{x_i | x_{B,i}^{\text{pri}}\} = \mathcal{W}(x_{B,i}^{\text{pri}}, \sigma_{B,\text{pri}}^2, s)$
8: $\sigma_{B,\text{post}}^2 = \frac{1}{L} \sum_{i=1}^{L} \mathrm{Var}\{x_i | x_{B,i}^{\text{pri}}\}$
9: $\sigma_{A,\text{pri}}^2 = \sigma_{B,\text{ext}}^2 = \left( \frac{1}{\sigma_{B,\text{post}}^2} - \frac{1}{\sigma_{B,\text{pri}}^2} \right)^{-1}$
10: $x_A^{\text{pri}} = x_B^{\text{ext}} = \sigma_{B,\text{ext}}^2 \left( \frac{x_B^{\text{post}}}{\sigma_{B,\text{post}}^2} - \frac{x_B^{\text{pri}}}{\sigma_{B,\text{pri}}^2} \right)$
11: }
12: $\hat{x} = \mathcal{Q}_{\mathcal{C}_0}(\hat{x})$

---

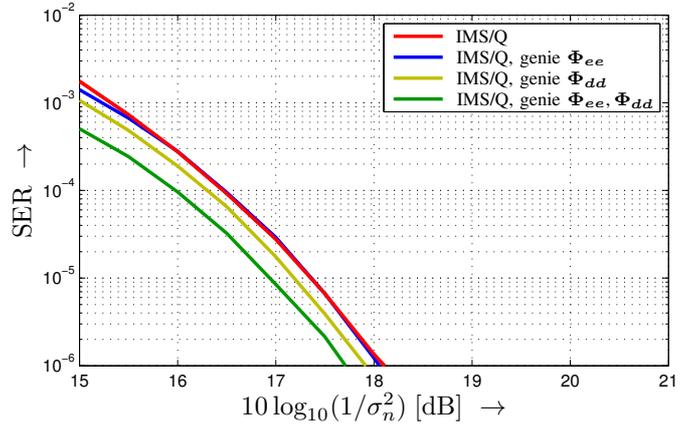

Fig. 3: SER of the proposed algorithm IMS over the noise level $1/\sigma_n^2$ in dB. $L = 258$, $K = 129$, $s = 20$, $\mathcal{C} = \{-1, +1\}$. SVD-based measurement matrix.

## IV. NUMERICAL RESULTS

In this section, the performance of the proposed algorithms is compared to the one of established algorithms. When dealing with discrete symbols, the measure of interest is the <u>s</u>ymbol <u>e</u>rror <u>r</u>ate (SER) (SER $= \frac{1}{L} \sum_{i=1}^{L} \Pr\{\hat{x}_i \neq x_i\}$), which is achieved for a certain noise level. Unless mentioned otherwise, the numerical evaluations are performed for $L = 258$, $K = 129$, $s = 20$. The measurement matrix is constructed as normalized random part of an orthogonal matrix of appropriate size which is obtained by an SVD from a random Gaussian matrix. In order to ensure convergence, all algorithms performed 50 iterations unless mentioned otherwise.

In Fig. 3, the achievable performance of IMS/Q (red line) is compared to the result which would be possible if the true actual variances of the estimation errors were known. In this case, the actual (true) error covariance matrix $\mathbf{\Phi}_{dd}$ and/or $\mathbf{\Phi}_{ee}$ is plugged in instead of the estimated one. If only the error covariance $\mathbf{\Phi}_{ee}$ after the first step is perfectly known but $\mathbf{\Phi}_{dd}$ is still estimated (blue), no significant improvement can be observed compared to IMS/Q without genie-aided knowledge. The estimation in this step is performed jointly, and the variances $\sigma_{e,i}^2$ of the elements of the sparse vector do not differ very much. If the actual variances after the second (non-linear) step were known (yellow), the performance could be improved slightly. In this case, the individual estimation of the elements leads to quite different reliabilities of the elements, and there are no averaging effects due to joint processing. However, even if both $\mathbf{\Phi}_{dd}$ as well as $\mathbf{\Phi}_{ee}$ were known exactly (green), only a small gain compared to IMS/Q with estimated error covariance matrices could be achieved. Thus, the loss of IMS/Q due to wrong error covariance estimates is less than $0.5\,\text{dB}$.

In Fig. 4 (top), the performance of IMS/Q and the adapted TSR/Q is compared to the one of state-of-the-art algorithms, such as IHT, IST, and OMP with successive quantization [11], respectively, as well as for OMP with subsequent vector quantization (OMP/SD) [11]. The number of iterations for OMP/Q has been numerically optimized with respect to the noise level. It ranges from 23 for $1/\sigma_n^2 \,\widehat{=}\, 15\,\text{dB}$ to 33 for $1/\sigma_n^2 \,\widehat{=}\, 21\,\text{dB}$ (cf. [11]). In the case of OMP/SD, the OMP performs 30 iterations. TSR/Q stops if $\sigma_{A,\text{pri}}^2$ is smaller than MATLAB precision or if the maximum number of iterations ($= 50$) is reached.

OMP with vector quantization (OMP/SD, black) shows the best performance of all state-of-the-art algorithms. The generalized TSR algorithm (TSR/Q, green) performs equal as OMP/SD in the case of high noise levels, but does not show the error floor for low noise powers. The proposed IMS/Q algorithm clearly outperforms all state-of-the-art algorithms by at least $2\,\mathrm{dB}$ for low noise levels, and tolerates $0.7\,\mathrm{dB}$ more noise power than TSR/Q. Also the so-called generalized approximate message passing (GAMP/Q, purple), which also employs soft feedback, is outperformed by IMS/Q.

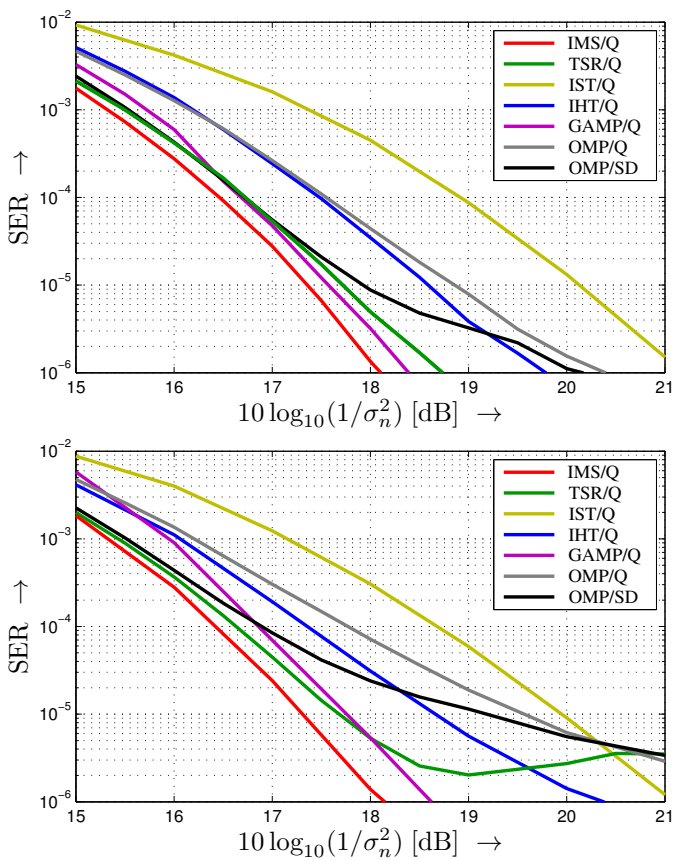

Fig. 4: SER of the proposed algorithms over the noise level $1/\sigma_n^2$ in dB. $L = 258$, $K = 129$ (upper part), and $L = 150$, $K = 100$ (lower part). $s = 20$, $\mathcal{C} = \{-1, +1\}$. SVD-based measurement matrix.

The same conclusion also holds in the case of $L = 150$ and $K = 100$, as can be seen in Fig. 4 (bottom). Compared to the previous setup, the relative sparsity (i.e., $s/L$) is larger, but also the relative number of measurements is increased, i.e., $K/L = 2/3$ (instead of $K/L = 1/2$). As before, IMS/Q outperforms all other algorithms, whereas GAMP/Q and TSR/Q are the best alternatives. Note that TSR/Q diverges in same rare cases, which causes the flattening at low noise levels.

## V. CONCLUSION

In this paper, a new algorithm IMS/Q has been proposed which solves the discrete compressed sensing problem in a very intuitive yet optimum way given the signal model and the chosen optimization criteria. Numerical results showed that this new algorithm clearly outperforms established algorithms. A convergence analysis of IMS/Q is the subject of current work and will be presented in a future publication. Furthermore, the turbo-principle-based TSR algorithm has been adapted to the discrete setup. It has been generalized and the procedure has been simplified without any change in performance.

## REFERENCES

[1] H. Zhu, G.B. Giannakis. Exploiting Sparse User Activity in Multiuser Detection. *IEEE Tr. Comm.*, pp. 454–465, Feb. 2011.
[2] R.F.H. Fischer, F. Wäckerle. Peak-to-Average Power Ratio Reduction in OFDM via Sparse Signals: Transmitter-Side Tone Reservation vs. Receiver-Side Compressed Sensing. *Proc. Int. OFDM Workshop*, Essen, Germany, Aug. 2012.
[3] A. Ens, A. Yousaf, T. Ostertag L.M. Reindl. Optimized Sinus Wave Generation with Compressed Sensing for Radar Applications. *Proc. CoSeRa*, Bonn, Germany, Sept. 2013.
[4] P. Dymarski, R. Romaniuk. Sparse Signal Modeling in a Scalable CELP Coder. *Proc. EUSIPCO*, Marrakech, Morocco, Sep. 2013.
[5] R. Fay. Introducing the Counter Mode of Operation to Compressed Sensing Based Encryption. *Inf. Proc. Letters*, pp. 279–283, Apr. 2016.
[6] D.L. Donoho. Compressed Sensing. *IEEE Tr. Inf. Theory*, pp. 1289–1306, Apr. 2006.
[7] Y.C. Pati, R. Rezaiifar, P.S. Krishnaprasad. Orthogonal Matching Pursuit: Recursive Function Approximation with Applications to Wavelet Decomposition. *Proc. Asilomar Conf.*, pp. 40–44, Nov. 1993.
[8] T. Blumensath, M.E. Davis. Iterative Thresholding for Sparse Approximations. *Journal of Fourier Analysis and Appl.*, pp. 629–654, Dec. 2008.
[9] I. Daubechies, M. Fornasier, I. Loris. Accelerated Projected Gradient Method for Linear Inverse Problems with Sparsity Constraints. *Journal of Fourier Analysis and Appl.*, pp. 764–792, Dec. 2008.
[10] G.L. Nehmhauser, L.A. Wolsey. *Integer and Combinatorial Optimization*, John Wiley & Sons, New York, 1988.
[11] S. Sparrer, R.F.H. Fischer. Adapting Compressed Sensing Algorithms to Discrete Sparse Signals. *Proc. Workshop on Smart Antennas (WSA) 2014*, Erlangen, Germany, Mar. 2014.
[12] E. Agrell, T. Eriksson, A. Vardy, K. Zeger. Closest Point Search in Lattices. *IEEE Tr. Inf. Theory*, pp. 2201–2214, Aug. 2002.
[13] R.G. Baraniuk, V. Cevher, M.F. Duarte, C. Hedge. Model-Based Compressive Sensing. *IEEE Trans. Inf. Theory*, pp.1982–2001, Apr. 2010.
[14] S. Sparrer, R.F.H. Fischer. Soft-Feedback OMP for the Recovery of Discrete-Valued Sparse Signals. *EUSIPCO*, Nice, France, Aug. 2015.
[15] S. Sparrer, R.F.H. Fischer. An MMSE-Based Version of OMP for the Recovery of Discrete-Valued Sparse Signals. *Electronics Letters*, pp. 75–77, Jan. 2016.
[16] D.L. Donoho, A. Maleki, A. Montanari. Message Passing Algorithms for Compressed Sensing: I. Motivation and Construction. *Proc. Information Theory Workshop (ITW)*, Cairo, Egypt, Jan. 2010.
[17] M. Bayati, A. Montanari. The Dynamics of Message Passing on Dense Graphs, with Application to Compressed Sensing. *IEEE Tr. Inf. Theory*, pp. 764–785, Feb. 2011.
[18] F.R. Kschischang, B.J. Frey, H.-A. Loeliger. Factor Graphs and the Sum-Product Algorithm. *IEEE Tr. Inf. Theory*, pp. 498–519, Feb. 2001.
[19] W. Dai, O. Milenkovic. Sparse Weighted Euclidean Superimposed Coding for Integer Compressed Sensing. *Proc. Conf. on Inf. Systems and Sciences (CISS)*, pp. 470–475, Princeton, New Jersey, USA , March 2008.
[20] W. Dai, O. Milenkovic. Weighted Euclidean Superimposed Codes for Integer Compressed Sensing. *Proc. Information Theory Workshop (ITW)*, pp.124–128, Porto, Portugal, May 2008.
[21] J. Ma, X. Yuan, L. Ping. Turbo Compressed Sensing with Partial DFT Sensing Matrix. *IEEE Signal Proc. Letters*, pp. 158–161, Feb. 2015.
[22] T. Frey, M. Reinhardt. Signal Estimation for Interference Cancellation and Decision Feedback Equalization. *Proc. IEEE Vehicular Techn. Conf.*, pp. 113–121, Phoenix, USA, May 1997.
[23] A. Lampe, J.B. Huber. On Improved Multiuser Detection with Iterated Soft Decision Interference Cancellation. *Proc. Comm. Theory Mini-Conf. at GLOBECOM*, pp. 172–176, June 1999.
[24] R.R. Müller, J.B. Huber. Iterated Soft-Decision Interference Cancellation for CDMA. *Broadband Wireless Comm.*, eds. M. Luise and S. Pupolin, pp. 110–115, Springer London, 1998.
[25] S.M. Kay. *Fundamentals of Statistical Signal Processing: I. Estimation Theory*, Prentice-Hall Inc., Upper Saddle River, NJ, USA, 1993.
[26] A. Engelhart, W.G. Teich, J. Linder. Complex Valued Signal Estimation for Interference Cancellation Schemes. *Tech. Rep. ITUU-TR-1998/04*, Dept. of Inf. Tech., Uni Ulm, Dec. 1998.
[27] J. Ma, X. Yuan, L. Ping. On the Performance of Turbo Signal Recovery with Partial DFT Sensing Matrices. *IEEE Signal Proc. Letters*, pp. 1580–1584, Oct. 2015.
[28] T. Lie, C.-K. Wen, S. Jin, X. You. Generalized Turbo Signal Recovery for Nonlinear Measurements and Orthogonal Sensing Matrices. *arXiv*, 1512.04833v4, 5. May 2016.
[29] C. Berrou, A. Glavieux. Near Shannon Limit Error-Correcting Coding and Decoding: Turbo-Codes. *Proc. IEEE ICC*, pp. 1064–1070, Geneva, Switzerland, May 1993.
[30] Q. Guo, D.D. Huang. A Concise Representation for the Soft-In Soft-Out LMMSE Detector. *IEEE Comm. Letters*, pp. 566–568, May 2011.